\documentclass[aps,prl,twocolumn,superscriptaddress,amsfonts,amssymb,amsmath,showpacs,showkeys,fleqn]{revtex4}
\usepackage{graphics}
\usepackage{epsfig}

\begin{document}

\title{Solid oxygen as converter for the production of ultra-cold neutrons}

\author{A.~Frei}\email[Corresponding author: ]{Andreas.Frei@tum.de}
\author{F.~B\"ohle}
\author{R.~Bozhanova}
\author{E.~Gutsmiedl}
\author{T.~Huber}
\affiliation{Physik Department, Technische Universit\"at M\"unchen, James-Franck-Str.~1, D-85748 Garching}
\author{J.~Klenke}
\affiliation{Forschungsneutronenquelle Heinz Maier-Leibnitz (FRM~II), Technische Universit\"at M\"unchen, Lichtenbergstr.~1, D-85748 Garching}
\author{S.~Paul}
\author{S.~Wlokka}
\affiliation{Physik Department, Technische Universit\"at M\"unchen, James-Franck-Str.~1, D-85748 Garching}

\begin{abstract}
We have investigated solid oxygen as a converter material for the production of ultra-cold neutrons. In a first series of experiments the crystal preparation was examined. An optically semi-transparent solid $\alpha$-oxygen crystal has been prepared. In a second series of experiments such a crystal prepared indentically as in the first series of experiments has been exposed to the cold neutron flux of the MEPHISTO beam line of the FRM II. Ultra-cold neutrons produced inside the oxygen crystal have been extracted and the count rates have been measured at different converter temperatures. The results of these measurements give a clear signal of the superthermal UCN production mechanism in $\alpha$-oxygen. The mean free loss length of UCN inside the crystal at a temperature of 5\,K was determined to be in the order of $20\,\mathrm{cm}$.
\end{abstract}



\maketitle

Precision experiments with ultra-cold neutrons (UCN), such as the search for a possible electric dipole moment (EDM) of the neutron or the measurement of the lifetime $\tau_{\mathrm n}$ of the free neutron, require high UCN densities. Stronger UCN sources are presently developed \cite{trinks2000, masuda2002, baker2003, saunders2004, frei2007, korobkina2007, zimmer2007, anghel2009, serebrov2009}, based on the principle of superthermal UCN production \cite{golub1975, golub1983} using cryo-converters made of solid deuterium, superfluid helium or solid oxygen. Until now promising results have been achieved with converters made of deuterium or helium. Another very promising candidate is solid oxygen, where UCN can be produced via phonon and magnon excitations in the crystal lattice. It has a large neutron-magnon inelastic scattering cross section and a small nuclear absorption cross section, and offers therefore the possibility to install large (several dm) converters at UCN sources with high ultra-cold neutron production rates. Since oxygen is antiferromagnetic and magnons can be excited only at the $\alpha$-phase ($T<23.9\,\mathrm{K}$) a solid oxygen crystal has to be grown. It should also be optically transparent to minimize losses during the extraction of UCN out of the converter.

In solid oxygen at low temperatures the thermal excitations of the lattice freeze out and long range antiferromagnetic ordering is established, leading to a two-dimensional antiferromagnetic structure \cite{stephens1986}. In addition to phonon excitations also magnetic spin wave excitations (magnons) contribute to the neutron scattering. This supplementary magnetic scattering of neutrons might be a strong down-conversion channel, which would enhance the production of ultra-cold neutrons. Theoretically this process was considered for the first time by \cite{liu2004}. Experimentally the neutron scattering cross sections and UCN production cross sections of solid $\alpha$-oxygen have been investigated by different groups, e.g. \cite{atchison2009, frei2010}, concluding that the UCN production rate should increase by cooling down the solid oxygen crystal to temperatures below 20\,K. But direct measurements of UCN count rates from such crystals conducted recently could not confirm the expected results \cite{salvat2009, atchison2010}. On the contrary the measured UCN count rates did not increase when lowering the crystal temperature, or even decreased. The reason for this is still unknown and most assumptions hold the losses due to UCN-extraction out of the oxygen crystal responsible for the measured effects. We report here on a series of measurements, on the one hand investigating the optical properties of an oxygen crystal, and on the other hand measuring the temperature dependence of count rates of UCN produced by an oxygen converter.

Ultra-cold neutrons have energies $E<300\,\mathrm{neV}$, corresponding to de Broglie wavelengths $\lambda>50\,\mathrm{nm}$. Thus the wavelength spectrum of ultra-cold neutrons with the lowest energies is close to the spectrum of visible light. Therefore the optical properties of an oxygen crystal are important. Optical transparency of the crystal is a prerequisite to minimize losses occuring during the extraction of UCN out of the crystal.

\begin{table*}
\begin{tabular}{lccccc}\hline\hline
Phase & Color & Pressure & Temperature & Structure & Symmetry \\ \hline
$\alpha$-phase & light blue & vapor & $<23.9\,\mathrm{K}$ & monoclinic & $C2/m$ \\
$\beta$-phase & blue & vapor & $23.9\,\mathrm{K}-43.8\,\mathrm{K}$ & rhombohedral & $R\overline{3}m$ \\
$\gamma$-phase & faint blue & vapor & $43.8\,\mathrm{K}-54.4\,\mathrm{K}$ & cubic & $Pm3n$ \\
$\delta$-phase & orange & $9.6\,\mathrm{GPa}-10\,\mathrm{GPa}$ & --- & orthorhombic & $Fmmm$ \\
$\epsilon$-phase & red & $10\,\mathrm{GPa}-96\,\mathrm{GPa}$ & --- & monoclinic & $A2/m$ \\
$\zeta$-phase & metallic & $>96\,\mathrm{GPa}$ & --- & metallic & --- \\ \hline\hline
\end{tabular}
\caption{Solid state phases of oxygen \cite{freiman2004}.}
\label{tab1}
\end{table*}

\begin{figure}
\begin{center}
\includegraphics[width=0.45\textwidth,keepaspectratio]{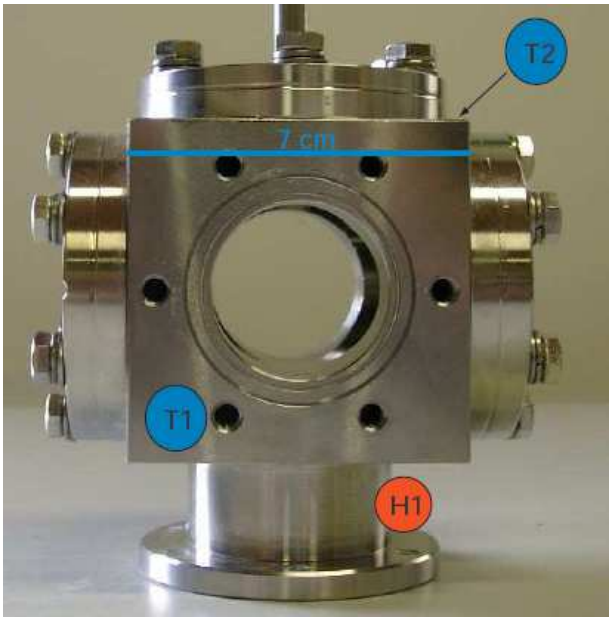}
\caption{Sample cell (dismounted) for optical investigation of the oxygen crystal. Material: Aluminium. Down side can be connected to a cold head. H1 shows the position of a $70\,\mathrm{W}$ heater. T1 and T2 are the positions of temperature sensors. Gas inlet from top. Not shown: Glass windows on the front and rear side for optical inspection.}
\label{fig2}
\end{center}
\end{figure}

There are six solid state phases of oxygen known to exist, but only three of them occur at vapor pressure \cite{freiman2004}. An overview of the characteristics of these phases is given in Tab.~\ref{tab1}. With our setup no pressure could be applied to the crystal, so only the $\alpha$-, $\beta$- and $\gamma$-phases occured during our measurements. In our experiment the procedure for generating an $\alpha$-oxygen crystal is the following: Cooling of the sample cell to $T\sim80\,\mathrm{K}$, liquification of inflowing oxygen gas and further cooling of the sample cell down to $T\approx10\,\mathrm{K}$ at vapor pressure conditions. Hereby the transitions $\mathrm{liquid}\rightarrow\gamma$, $\gamma\rightarrow\beta$ and $\beta\rightarrow\alpha$ occur during the cooling process. Before measuring the UCN production at these different phases the oxygen crystal was optically inspected, and the cooldown rates especially at the phase transitions were optimized. Therefore a sample cell ($70\times70\times70\,\mathrm{mm}$) made of aluminium is connected to a 4K-Cold-Head (type RDK-415D of Sumitomo Heavy Industries ltd.) with a power of $1.5\,\mathrm{W}$ at $4\,\mathrm{K}$ at its bottom side. The usable volume for the oxygen inside the cell is $35\times35\times35\,\mathrm{mm}$, and the inner walls of the cell (exept for the extraction window) are coated with Ni. At the bottom of the cell a $70\,\mathrm{W}$ heater is installed to adjust the temperature, which is measured by two temperature sensors (type Cernox CX-1050-AA of LakeShore Cryotonics Inc.) at the bottom and at the top of the cell. As gas inlet a stainless steel tube (outer diameter 8\,mm, wall thickness 1\,mm) is connected to the top. At the front and rear side two glass windows are mounted to the cell for optical inspection of the oxygen volume. Fig. \ref{fig2} shows a picture of the sample cell.

In a first series of experiments the cooldown rates especially at the phase transitions were optimized, and the oxygen crystal was optically inspected. At the transition $\mathrm{liquid}\rightarrow\gamma$, which occurs at $54.4\,\mathrm{K}$ at vapor pressure \cite{young1987}, the cooling rate had no influence on the transparency of the resulting $\gamma$-crystal. Hereby the cooling rates were varied in the range of $1.0-0.1\,\mathrm{K/h}$. In any case this crystal had a faint blue color and was optically transparent.

The $\gamma\rightarrow\beta$ phase transition is a polymorphic transformation from the cubic $\gamma$-phase to the rhombohedral $\beta$-phase at a temperature of $43.8\,\mathrm{K}$ \cite{cowen1976} with a heat of transition of $\Delta H\approx742\,\mathrm{J/mol}$ \cite{fagerstroem1969} and a volume jump of $\Delta V\approx1.08\,\mathrm{cm^3/mol}$ \cite{jahnke1967}. Due to these circumstances this transition is critical concerning the optical transparency of the resulting $\beta$-crystal. In our experiments we found that only at cooling rates of below $10\,\mathrm{mK/h}$ the resulting $\beta$-crystal was optically cloudy with a blue color. At faster cooling rates the resulting $\beta$-crystal was completely black and optically opaque.

\begin{figure}
\begin{center}
\includegraphics[width=0.45\textwidth,keepaspectratio]{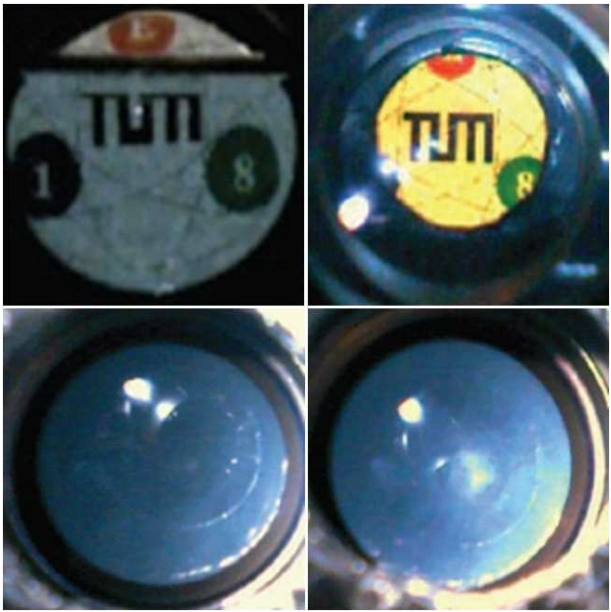}
\caption{Pictures of the oxygen crystal at different phases. Top left: Liquid oxygen in the sample cell (partially filled). Top right: Transparent $\gamma$-oxygen (sample cell completely filled). Bottom left: Cloudy $\beta$-oxygen. Bottom right: Cloudy $\alpha$-oxygen.}
\label{fig3}
\end{center}
\end{figure}

The $\beta\rightarrow\alpha$ phase transition occurs at a temperature of $23.9\,\mathrm{K}$ \cite{freiman2004}. Questions like this transition being of second order or not, on heat of transition or a volume jump is under dispute. While a heat of transition and a volume jump were measured by several groups \cite{eucken1916, clusius1929, giauque1929, ancsin1975}, these results have been questioned by several other groups \cite{fagerstroem1969, muijlwijk1969}. In our experiment the cooling rate was variied in the range of $1.0-0.1\,\mathrm{K/h}$ and had no influence on the optical properties of the resulting $\alpha$-phase crystal, which showed the same color and transparency as the $\beta$-crystal. In Fig. \ref{fig3} pictures of the different crystals described above can be seen.

In a second series of experiments the sample cell was placed in the cold neutron beam line MEPHISTO of the FRM~II, and exposed to a cold neutron flux of $\Phi=5.25\cdot10^9\,\mathrm{cm^{-2}s^{-1}}$. The energy distribution of the cold neutrons has almost Maxwellian shape, with a characteristic temperature of $T=40\,\mathrm{K}$ \cite{frm2jahresbericht2007}. The glass windows of the sample cell were replaced by thin aluminium windows (thickness $500\,\mathrm{\mu m}$, diameter $35\,\mathrm{mm}$, coated with Ni on the inner side) to allow transparency of incoming and outgoing neutrons. UCN produced were extracted horizontally and perpendicular to the incoming beam through an aluminium window (thickness $200\,\mu\mathrm{m}$, diameter $35\,\mathrm{mm}$) and guided through an electropolished stainless steel tube (inner diameter $66\,\mathrm{mm}$, total length $\sim3\,\mathrm{m}$) to a UCN detector of the CASCADE type \cite{klein2001}. The UCN extraction tube was bend by $90^\circ$ in the horizontal plane $50\,\mathrm{cm}$ away from the sample cell, then had a straight section of $2\,\mathrm{m}$, and was bent again by $90^\circ$ vertically downwards towards the detector. As an option a stainless steel foil (thickness $100\,\mu\mathrm{m}$) could be inserted into the UCN extraction tube. By this only neutrons with energies $>190\,\mathrm{neV}$ (corresponding to the Fermi-Potential of the foil material \cite{ignatovich1990, golub1991}) can penetrate the foil and are counted in the detector, while all UCN with energies $<190\,\mathrm{neV}$ are reflected from the foil and don't reach the detector. By substracting count rates of measurements with the inserted foil from count rates without foil, the effective UCN count rate with UCN energies $<190\,\mathrm{neV}$ can be obtained. Hereby the count rates with foil were 6-7 times smaller than count rates without foil, which is an indication that most of the UCN with energies $>190\,\mathrm{neV}$ could not reach the detector and the stainless steel foil anyhow, as they have to pass two $90^\circ$ bends before they reach the foil and are mainly lost in these bends. Before substraction of the count rates with foil from the count rates without foil, the first ones have been corrected for absorption and back-reflection of UCN from the foil. These correction values, which contribute in absolute numbers only to a very small fraction, were determined by Monte Carlo simulations of the transport efficiency of UCN with different energies from the converter to the detector. Also background count rates of $0.1\,\mathrm{s^{-1}}$ with an empty converter cell have been observed.

\begin{figure}
\begin{center}
\includegraphics[width=0.45\textwidth,keepaspectratio]{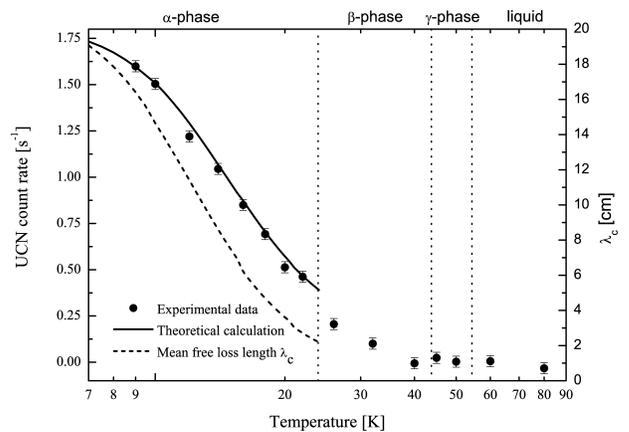}
\caption{Circles: Measured UCN count rates produced by a solid oxygen converter, depending on its temperature and solid state phase. Error bars indicate statistical uncertainties. Solid line: Calculated UCN count rate (see text for details). Dashed line: Mean free loss length $\lambda_c$ (axis to the right) in the oxygen converter used for the calculations.}
\label{fig4}
\end{center}
\end{figure}

In Fig.~\ref{fig4} the resulting UCN count rates measured at different temperatures of the oxygen converter are depicted. The UCN count rate increases with decreasing temperature. This trend is a clear indication of the theoretically expected UCN production and extraction mechanism via the superthermal principle of detailed balance. To compare quantitatively the measured UCN count rates with theory to following considerations have been taken into account: In the converter volume (subscript $c$ in formulas) an equilibrium between UCN production on the one hand, and UCN flow out of the converter plus losses inside the converter on the other hand will build up, which can be described with (see also \cite{trinks2000})
\begin{eqnarray}
NP\Delta EV_c & = & \Phi_\mathrm{out}A_\mathrm{out}+\rho_cV_c/\tau_c, \\
\Phi_\mathrm{out} & = & \rho_c\bar{v}_c/4,\\
\tau_c^{-1} & = & \tau_\beta^{-1}+\tau_\mathrm{abs}^{-1}+\tau_\mathrm{w}^{-1}+\tau_\mathrm{up}^{-1}+\tau_\mathrm{x}^{-1}.\label{eq3}
\end{eqnarray}
Here $N$ is the particle density of the converter, $P\Delta E$ the production rate per O-atom into the energy interval $\Delta E$, $V_c$ the converter volume, $\Phi_\mathrm{out}$ the UCN flux density through the area $A_\mathrm{out}$ out of the converter, $\bar{v}_c$ the UCN mean velocity, $\rho_c$ the UCN density and $\tau_c$ their effective storage time in the converter. Hereby it was assumed, that there is no flow of already extracted UCN back into the converter, which means that $\Phi_\mathrm{out}A_\mathrm{out}$ is proportional to the count rate at the detector. The production rate $P=\int\frac{d\sigma}{dE_idE_f}\Phi(E_i)dE_idE_f$ was calculated using the incoherent approximation according to \cite{turchin1965} for single lattice excitations, where $E_{i,f}$ denotes the initial and final neutron energy. Input parameters in this calculation are the density of states of $\alpha$-O$_2$ from \cite{frei2010} and the scattering cross section $\sigma=12.1\,\mathrm{barn}$ \cite{liu2004, sears1992} (nuclear and magnetic scattering). In this formalism production and upscattering of already produced UCN are calculated simultaneously. The calculated UCN production rate shows only slight variations with changing converter temperature from (5-24)$\,$K between (1.9-2.3)$\,\mathrm{cm^{-3}s^{-1}}$. $\tau_c$ can be calculated from all the loss mechanisms which occure: $\beta$-decay $\tau_\beta=885.7\,\mathrm{s}$ \cite{nakamura2010}; absorption $\tau_\mathrm{abs}^{-1}=N\bar{v}_c\sigma_\mathrm{abs,ucn}=Nv_\mathrm{th}\sigma_\mathrm{abs,th}$ with $N=5.8\cdot10^{22}\,\mathrm{cm^{-3}}$ \cite{roder1978}, $v_\mathrm{th}=2200\,\mathrm{m/s}$ and $\sigma_\mathrm{abs,th}=1.9\cdot10^{-28}\,\mathrm{cm^2}$ \cite{sears1992}, leading to $\tau_\mathrm{abs}=0.8\,\mathrm{s}$; losses due to wall collisions $\tau_\mathrm{w}^{-1}=(\bar{v}_c/s_c)\mu$ with $\bar{v}_c=4.2\,\mathrm{m/s}$, the mean free path between wall collisions $s_c=4V_c/A_c=2.3\,\mathrm{cm}$ and the loss probability per wall collision $\mu=1\cdot10^{-3}$ \cite{golub1991}, leading to $\tau_\mathrm{w}=5.5\,\mathrm{s}$. Monte Carlo simulations of the transport efficiency for UCN extracted from the converter travelling to the detector have been performed for the used neutron guide geometry, leading to a mean value (averaged over the UCN energy and for the case that no foil was inserted in the UCN guide) of $4.8\%$.

To fit the calculated values to the experimental data two scaling parameters have been introduced: A temperature independent parameter $\tau_\mathrm{x}$ in Eq. (\ref{eq3}), accounting for any non quantitatively known losses as e.g. the crystal quality, and a scaling parameter $\kappa$, which is multiplied to the scattering cross section $\sigma$. While $\tau_\mathrm{x}$ only scales linearly the calculated to the measured UCN count rate, the steepness of the count rate curve is affected by $\kappa$. The best fit with these two parameters of the theoretical count rate curve to the measured data resulted in $\tau_\mathrm{x}=56\,\mathrm{ms}$ and $\kappa=2.84$, where the uncertainty of the fit of each of the parameters is $15\%$. The measured UCN count rates, together with the best theoretical fit, and the hereby deduced effective mean free loss length $\lambda_c(T)=\bar{v}_c\tau_c(T)$ is depicted in Fig.~\ref{fig4}. As can be seen there, the mean free loss length becomes $\lambda_c(5\,\mathrm{K})\sim20\,\mathrm{cm}$. The uncertainty of $\lambda_c$ hereby is approximately $50\%$, because variations by $15\%$ in the fit parameters result in much larger variations in the saturation value of $\lambda_c$.

In summary, we have developed a method to produce an optically semi-transparent solid oxygen crystal in its $\alpha$-phase, which is a prerequisite to minimize extraction losses of UCN out of the crystal. The solid oxygen converter has been exposed to a cold neutron beam, and the UCN count rates have been measured. The measured temperature dependence of the UCN count rate can be explained with theoretical models of the superthermal UCN production mechanism. The usable converter size was determined to be in the order of $20\,\mathrm{cm}$, a value that is significantly smaller as the theoretical neutron absorption length of $380\,\mathrm{cm}$ \cite{liu2004}. This discrepancy motivates further measurements and improvements in this field of research.

This work was supported by the cluster of excellence ''Origin and Structure of the Universe'' (Exc 153) and by the Maier-Leibnitz-Laboratorium (MLL) of the Ludwig-Maximilians-Universit\"at (LMU) and the Technische Universit\"at M\"unchen (TUM). We thank T.~Deuschle and H.~Ruhland for their help during the experiments.

\end{document}